\documentstyle[prb,aps,epsfig]{revtex}

\def\E{{\rm e}}

\begin{document}
\draft

\title{Classical diffusion of $N$ interacting particles \\ in one
dimension: General results and asymptotic laws}

\author{Claude Aslangul\footnote
  {{\bf e-mail:} aslangul@gps.jussieu.fr}}
\address{Groupe de Physique des Solides, Laboratoire associ\'e au
CNRS (UMR 75-88), \\ Universit\'es Paris 7 \&
Paris 6, Tour 23, Place Jussieu, 75251 Paris Cedex 05, France}

\maketitle
\begin{abstract}
I consider the coupled one-dimensional diffusion of a cluster of $N$
classical particles with contact
repulsion. General expressions are given for the probability
distributions, allowing to obtain the transport coefficients.
In the limit of large $N$, and within a gaussian approximation,
the diffusion constant is found to behave as
$N^{-1}$ for the central particle and as $(\ln N)^{-1}$ for the edge
ones. Absolute correlations between the edge particles {\it increase}
as $(\ln N)^{2}$.
The asymptotic one-body distribution is obtained and discussed in
relation of the statistics of extreme events.
\end{abstract}

\vspace*{0.6cm}

\section{Introduction}
Low dimensionality, geometrical constraints and interactions between
classically diffusing particles are expected to modify transport
coefficients and/or the nature of the asymptotic regimes. As a
trivial example, a brownian particle subjected to a purely reflecting wall
has a anomalous drift, its average position increasing as $t^{1/2}$,
(which corresponds to a vanishing velocity)
whereas the centered second moment has a normal diffusive
spreading  with a lowered diffusion
constant as compared to the diffusion without barrier. In the case of two
particles with a contact repulsion, each of them plays for the other
the role of a {\it fluctuating} boundary condition which affects both the
transport coefficients linked to the average position and the mean square
deviation.
The aim of this letter is to discuss such questions in the general
case of $N$ mutually interacting particles with a hard-core
interaction.

Classical diffusion with interactions does not seem to have drawn so much
attention. A notable exception is the so-called ``tracer problem''
defined as the diffusion of a tagged particle in an infinite sea of
other diffusing particles. The one-dimensional model was first solved by
Harris \cite{har65} and
discussed in subsequent papers \cite {vanB83,brum88}. The main result
is that the mean square dispersion of the position, $\Delta x^{2}$,
displays a subdiffusive behaviour, which originates from the fact that the
motion of the tagged particle is,
anywhere and at any time, hindered by all surrounding particles. In
one dimension Harris found that $\Delta x^{2}$ grows as $t^{1/2}$ at large
times; this implies that the typical distance travelled by the tracer at
time $t$
goes like $t^{1/4}$ instead of $t^{1/2}$ in the free case.
More recently, Derrida et al. \cite{der:muk} found several exact results
for the asymmetric simple
exclusion process \cite{lig85}; recent bibliography and other
results on such subject can be found in Mallick's thesis \cite{mal96}.

The problem here considered is frankly different, although it belongs
to the so-called single-file diffusion problem encountered in many
fields (one-dimensional hopping conductivity \cite{rich},
ion transport in biological membranes \cite{nener,sack},
channelling in zeolithes \cite{kukla}). At some initial
time, a compact cluster of $N$ pointlike particles is launched at the
origin of a one-dimensional space; each of them undergoes ordinary brownian
motion
but has a contact repulsive interaction with its neighbours. As a
consequence, the particles located at the egde of the cluster can move
freely
on one side and are subjected to a fluctuating boundary condition on
the other, whereas the particles inside the cluster are subjected to
such boundary conditions on either side.

The questions to be solved
are to find the (anomalous) drift of one particle and its diffusion
constant as a function of the position within the cluster and of the
number of $N$. In addition, two-particle correlations are worthy to be
analyzed, as well as the asymptotic one-body probability distributions.

%
\section{One-particle transport coefficients}
At the initial time, the $N$ particles are assumed to form a compact
cluster located at the origin $x=0$, each of them having the same
diffusion constant $D$ as all the others. Due to the contact repulsion, two
particles can never cross each other, so that order in space is
preserved at any time; this means that the $N$ coordinates $x_{i}$
can be labelled so that:
\begin{equation}
x_{1}\,<\,x_{2}\,<\, \ldots \,<\, x_{N} \qquad \forall t
\enspace.\label{xord}
\end{equation}
The solution of the diffusion equation for such an initial condition
is the following:
\begin{equation}
p(x_{1},x_{2},\ldots,x_{N};t) =
N!\prod_{n=1}^{N}\frac{\E^{-x_{i}^{2}/(4Dt)}}{\sqrt{4\pi Dt}}\
\prod_{n=1}^{N-1}
Y(x_{i+1}-x_{i}) \enspace,\label{PN}
\end{equation}
where $Y$ is the Heaviside unit step function ($Y(x)=1$ if $x>0$, $0$
otherwise). In a recent and important paper \cite {Rod98}, the general
formal solution of the same problem
with an arbitrary initial condition was given, using the reflection
principle.
>From eq. (\ref {PN}), one readily gets the reduced one-particle density
for the $n^{\rm th}$ particle of the cluster:
\begin{equation}
p_{n}^{(1)}(x;t)\,=\,
\frac{2^{1-N}N!}{(n-1)!(N-n)!}\left[1+\Phi\left
(\frac{x}{\sqrt{4Dt}}\right)\right]^{n-1}
\left[1-\Phi\left(\frac{x}{\sqrt{4Dt}}\right)\right]^{N-n}
\frac {\E^{-x^{2}/(4Dt)}}{\sqrt{4\pi Dt}}
\enspace.\label{pnN}
\end{equation}
In the last equation, $\Phi$ denotes the probability integral \cite
{Gr:Ry}, satisfying $\Phi(\pm\infty)=\pm 1$. The two factors $(1\pm\Phi)$
represent the steric effects on the
$n^{\it th}$ particle due to the other ones. Knowing $p_{n}^{(1)}$, it is
in principle possible to compute the
first few moments giving the average position and the mean square
dispersion for anyone of the particles. As a first result, one
immediately observes that any moment of the coordinate has a quite simple
variation in
time, at any time, not only in the final stage of the motion. Indeed,
using eq. (\ref {pnN}), it is readily seen that the the $k^{{\rm th}}$
moment $<x_{n}^{k}>$ increases at any time as $t^{k/2}$
since $(Dt)^{1/2}$ is the only lengthscale
of the pro\-blem. As a consequence, $\forall t$, the average coordinate
of the $n^{{\rm th}}$ particle $<x_{n}>(t)$ increases
as $t^{1/2}$ -- except for instance for the central particle of the cluster
when $N$ is
odd --, whereas the mean square displacement
$\Delta x_{n}^{2}\,\equiv\,<x_{n}^{2}>\,-\,<x_{n}>^{2}$ increases as $t$.
The drift, due to left-right symmetry breaking, is thus
always anomalous and the diffusion
is always normal. As a consequence, one can define, $\forall n$ and
$N$, the following transport coefficients $V_{1/2,\,n}$ and $D_{n}$:
\begin{equation}
<x_{n}>\,=\,V_{1/2,\,n}(N)\,t^{\frac{1}{2}}
\enspace,\label{x1n}
\end{equation}
\begin{equation}
\Delta x_{n}^{2}\,=\,2D_{n}(N)\,t\enspace,\label{deltax2n}
\end{equation}
It remains to find the functions $V_{1/2,\,n}(N)$ and  $D_{n}(N)$,
which incorporate the dependence of the transport coefficients upon
the number of particles. For $N=2$, one readily finds:
\begin{equation}
<x_{2}>\,=\,-\,<x_{1}>\,=\,\sqrt{\frac{2}{\pi}Dt}
\enspace,\label{x12}
\end{equation}
and
\begin{equation}
\Delta x_{n}^{2}\,=\,2(1-\frac{1}{\pi})\,t\enspace.\label{deltax22}
\end{equation}
In the case of a non-fluctuating (fixed) perfectly reflecting barrier, one has:
\begin{equation}
<x>\,=\,\sqrt{\frac{4}{\pi}Dt}
\qquad \Delta x^{2}\,=\,2(1-\frac{2}{\pi})\,t
\enspace.\label{fixe}
\end{equation}
Thus, for a cluster of two particles, each of them acting for the
other as a
fluctuating barrier, the drift is slowered and the
diffusion is enhanced as compared to a fixed barrier. These facts are
easily understood on physical grounds.

Unfortunately, it does not seem possible to write the exact
expressions of $V_{1/2,\,n}$ and $D_{n}$
in a closed, tractable form, starting
from eq. (\ref {pnN}). On the other hand, since it is worthy to analyze
the case of a large number $N\gg 1$ and since the factors involving
the $\Phi$'s functions
have rather sharp derivatives, especially when $N$ is large, it is
expected that a
gaussian approximation can indeed produce the correct large-$N$ variation of
the $V_{1/2,\,n}$ and  $D_{n}$.

Let us first consider one of the two particles located at one
extremity of the cluster, the right one for instance. From eq. (\ref
{pnN}), the one-body density can be written as ($u\,=\,x/\sqrt{4Dt}$):
\begin{equation}
p_{N}^{(1)}(x;t)\,=\,\frac{1}{\sqrt{4Dt}}\,\frac{d}{du}
\left[\frac{1+\Phi(u)}{2}\right]^{N}
\enspace.\label{pNN}
\end{equation}
This expression naturally has the form encountered in the statistics of
extreme
values \cite {gum}. When $N\gg 1$, this is a very sharp function with a
maximum $u_{0}$
defined by:
\begin{equation}
\frac{\E^{-u_{0}^{2}}}{u_{0}}\,\simeq\,\frac{2\sqrt{\pi}}{N}
\enspace.\label{u0}
\end{equation}
Making now a gaussian approximation for $p_{N}^{(1)}(x;t)$, one
readily finds, up to logarithmic corrections:
\begin{equation}
<x_{N}>\,=\,-\,<x_{1}>\,\simeq\,\sqrt{\ln\frac{N}{2\sqrt{\pi}}}\,\sqrt{4Dt}
\enspace.\label{x1Ngauss}
\end{equation}
Thus, for large $N$, the coefficient for the anomalous drift has a
logarithmic increase with respect to the number $N$ of particles of
the cluster:
\begin{equation}
V_{1/2\,N}(N)\,\propto\,(\ln N)^{1/2}
\enspace.\label{VNgauss}
\end{equation}
The fact that $V_{1/2\,N}$ increases with $N$ is evident on physical
grounds (all the ``inside" particles are pushing on those which are at
the edges), but this increase is extremely slow.
In addition, the same approximation yields:
\begin{equation}
\Delta x_{N}^{2}\,=\,\Delta x_{1}^{2}\,\simeq
\,\frac{\E^{2/3}}{(2\pi)^{1/3}\ln\frac{N}{2\sqrt{\pi}}}\,Dt
\enspace,\label{deltax2Ngauss}
\end{equation}
so that:
\begin{equation}
D_{1}(N)\,=\,D_{N}(N)\,\propto\,(\ln N)^{-1}
\enspace.\label{DNgauss}
\end{equation}
Although the diffusion is normal, the diffusion constant decreases
and tends toward $0$ for infinite $N$. In a pictorial way, the more
there are particles pushing on its back, the less quickly spreads any
one of the edge particles on either side of its average position,
which drifts proportionnally to $t^{1/2}$. Note that, from eqs. (\ref
{x1Ngauss})
and (\ref {deltax2Ngauss}), the relative fluctuations for the edge
particles behave as $(\ln N)^{-1}$; this extremely slow decrease of
fluctuations, as compared to $N^{-1/2}$ in ordinary cases, implies that
convergence toward a large-number law, if any, is quite poor.

The correctness of the gaussian approximation for the two first
moments was checked by numerically
computing the exact average position and exact mean square
displacement and by looking
at $<x_{N}>/[4Dt\,\ln(N/2\sqrt{\pi})]^{1/2}$ and
$[\Delta x_{N}^{2}/(4Dt)]\ln\frac{N}{2\sqrt{\pi}}$. Fig. 1 displays
the rather rapid
convergence of such quantities toward constants at large $N$,
confirming the validity of the gaussian approximation at least for the
two first moments.

Obviously, things go quite differently for the particle located at the
center of
the cluster, assuming $N$ to
be an odd number for simplicity. First, it does not move in the mean. Second,
it must
have a rather small diffusion constant as compared to the edge
particles, since it is strongly inhibited
by its numerous erratic partners on either side of it. Indeed, the gaussian
approximation yields:
\begin{equation}
\Delta x^{2}_{(N+1)/2}\,\simeq
\,\frac{\pi}{N}Dt\enspace.\label{x2center}
\end{equation}
This provides the large-$N$ dependence of the diffusion constant for
the central particle:
\begin{equation}
D_{(N+1)/2}(N)\,\propto\,\frac{1}{N}\enspace,\label{Dcentre}
\end{equation}
entailing that the fluctuations are now of the
order of $1/\sqrt{N}$.

Thus, in any case, the diffusion is normal, as contrasted to the Harris'
case for which $\Delta x^{2}\propto\,t^{1/2}$. Yet, note that in the
$N\rightarrow\infty$ limit, both $D_{N}$ and $D_{(N+1)/2}$ vanish,
which indicates a lowering of the dynamical exponent. The vanishing in
all cases
of the diffusion constants in the $N$-infinite limit signals the
onset of a subdiffusive regime in the finite concentration situation.
Considering the middle particle, which is surrounded by infinitely
many other, this is in conformity with Harris'
result. For the two edge particles, the marginal logarithmic
decrease of $D_{N}$ comes from the fact that the former still
face a free semi-infinite space to wander in.

Note that the scaling upon $N$ as described by eqs. (\ref{VNgauss}) and
(\ref{DNgauss}) are the same as those obtained in ref. \cite{boumez};
nevertheless, the asymptotic distribution law is not of the Gumbel
type (see below).
\section{Correlations}
Statistical correlations inside the cluster are also worthy to analyze.
As an example, let us consider the correlations between the two edge
particles. For the latter, the two-body probability density is easily found
from eq. (\ref{PN}) as the following:
\begin{equation}
p_{1\,N}^{(2)}(x_{1},x_{N};t)\,=\,	\frac{N(N-1)}{\pi\,2^{N}Dt}\,
\left[\Phi(u_{N})-\Phi(u_{1})\right]^{N-2}\E^{-(u_{1}^{2}+u_{N}^{2})}\,Y(u_{N}-u
_{1})
\enspace.\label{pRL2}
\end{equation}
Two-body correlations are most simply measured
by $C_{1\,N}\,=\,<x_{1}x_{N}>-<x_{1}><x_{N}>$; making again a gaussian
approximation
for the two-body density, this correlator has the following approximate
expression:
\begin{equation}
C_{1\,N}(t)\,\simeq \,  4\ln \frac{N}{2\sqrt{\pi}}\,Dt\enspace.\label{CRLN}
\end{equation}
Due to scaling in space, the normalized ratio $C_{1\,N}(t)/\Delta
x^{2}_{1}$ is a constant in time.
This constant turns out to be an {\it increasing}  function of the number
$N$ of particles; from eqs. (\ref{x1Ngauss}) and (\ref{deltax2Ngauss}), one
finds:
\begin{equation}
\frac{C_{1\,N}(t)}{\Delta x^{2}_{1}}\,\simeq \,4\,\left(\frac{2\pi}
{\E ^{2}}\right)^{1/3}\,\left[\ln
\frac{N}{2\sqrt{\pi}}\right]^{2}\enspace.\label{CRLNNorm}
\end{equation}
Thus, increasing the number of inner particles {\it enhances}, although
quite slowly, the correlations between the two edge particles. Far
from inducing some kind of screening effect, repeated numerous collisions
from inner particles enhance the statistical correlations between the
edge particles. In a pictorial way, it can be said that the former
act as ``virtual bosons" by going from one to the other edge particles;
the more they are, stronger is the effective (statistical) coupling.

\section{Asymptotic distribution laws}
Interestingly enough, it is also posible to obtain the asymptotic form
of the one-body distribution given by eq. (\ref {pnN}). For the right
particle ($n=N$), starting from eq. (\ref{pNN}), one easily finds, with still
$u=x/\sqrt{4Dt}\,>\,0$ :
\begin{equation}
p_{N}^{(1)}(x, t)\,\simeq \,\frac{N}{\sqrt{4\pi
Dt}}\,\left(1+\frac{1}{2u^{2}}\right)\,
{\rm exp}\left[-u^{2}\,-\frac{N}{2u\sqrt{\pi}}\,\E^{-u^{2}}\right]
\enspace.\label{p1Nas}
\end{equation}

The maximum occurs for $u\simeq u_{0}$ and the front is clearly
asymmetric around $u_{0}$ -- although the gaussian approximation, as shown
above, well
accounts for the large-$N$ dependence of the two first moments
(expectation value and fluctuations).
This asymmetry represents the pressure exerted by the inner
particles on the edge ones. For the left particle, one simply has
$p_{1}^{(1)}(x, t)=p_{N}^{(1)}(-x, t)$.
Fig. \ref{compppas} shows that the large-$N$ expression, eq. (\ref
{p1Nas}), reproduces
quite well the exact $p_{N}^{(1)}$ even for a moderately large value of
$N$. From eq. (\ref {p1Nas}), it is seen that $p_{N}^{(1)}(x, t)$ is
not exactly a Gumbel distribution; on the other hand, the rescaled
variable $u^{2} - \ln(N/2\sqrt{\pi})$ has, up to logarithmic
corrections, the same dependence upon $N$ as a true Gumbel variable
as far as the two first moments are concerned.

As constrasted, starting again from eq. (\ref {pnN}) for the central
particle ($n=(N+1)/2$), the asymptotic form of
the one-body density turns out to be simply the following normal law:
\begin{equation}
p_{(N+1)/2}^{(1)}(x, t)\,\simeq \,\frac{1}{\pi}\,\sqrt{\frac{N}{2 Dt}}\,
\E ^{-Nx^{2}/(2\pi Dt)}
\enspace,\label{p1Nasmil}
\end{equation}
in agreement with eq.(\ref {x2center}).

\section{Acknowlegdements}
I am indebted to Jean-Philippe Bouchaud for a helpful discussion on
the statistics of extremes.

%
%

\newpage
{\bf Figure Captions}
\begin{enumerate}
\item{}
\label{x(N)Dx2(N)}
Illustration of the asymptotic dependence upon $N$ of the
transport coefficients for the edge particles (average position,
eq. (\ref {x1Ngauss}) and mean square displacement, eq. (\ref
{deltax2Ngauss})).

\item{}
\label{compppas}
Comparison of the exact (solid line) and asymptotic distribution
(dashed line) functions, respectively given by eqs. (\ref {pnN}) and
(\ref {p1Nas}), for a cluster of $1000$ particles; the absissa is the
reduced variable $u=x/\sqrt{4Dt}$.

\end{enumerate}

\vskip-12pt
\begin{thebibliography}{99}

%
\bibitem{har65}
T. E. Harris J. Appl. Prob. {\bf 2}, 323, 1965.
%
\bibitem{vanB83}
H. van Beijeren, K. W. Kehr and K. Kutner Phys. Rev. A.
{\bf 28} ,5711, 1983.

%
\bibitem{brum88}
J. A. M. Brummelhuis and H. J. Hilhorst, J. Stat. Phys.
{\bf 53}, 249, 1988.
%
\bibitem {der:muk}
B. Derrida, E. Domany and D. Mukamel,
J. Stat. Phys., {\bf 69}, 1992, 667.
%
\bibitem {lig85}
T. M. Liggett,
{\it Interacting Particle Systems} (Springer Verlag, New York, 1985).
%
\bibitem {mal96}
K. Mallick,
{\it Syst{\`e}mes hors d'{\'{e}}quilibre : Quelques r\'{e}sultats
exacts} (Thesis, Universit\'e Paris-Sud, 1996).


%
\bibitem {rich}
P. M. Richards,
Phys. Rev. B, {\bf 16}, 1393, 1977.


%
\bibitem {nener}
E. Nener
Science, {\bf 256}, 498, 1992.

%
\bibitem {sack}
B. Sackman,
Science, {\bf 256}, 503, 1992.
%
\bibitem{kukla}
V. Kukla, J. Kornatowski, K. Demuth, I. Girnus, H. Pfeifer, L.
V. C. Rees, S. Schunk, K. K. Unger and J. K. K\"{a}rger,
Science, {\bf 272}, 702, 1996.

%
\bibitem {Rod98}
C. R\"{o}denbeck, J. K\"{a}rger and K. Hahn,
Phys. Rev. E, {\bf 57}, 4382,1998.

%
\bibitem{Gr:Ry}
I. S. Gradsteyn and I. M. Ryzhik,
{\it Table of Integrals, Series and Products}
(Academic Press, New York, 1980).

%
\bibitem{gum}
E. J. Gumbel
{\it Statistics of Extremes},
(Columbia University Press, New York \& London, 1958).

%
\bibitem {boumez}
J.-Ph. Bouchaud and M. Mezard,
J. de Phys. A, {\bf 30}, 7997,1997.



\end{thebibliography}
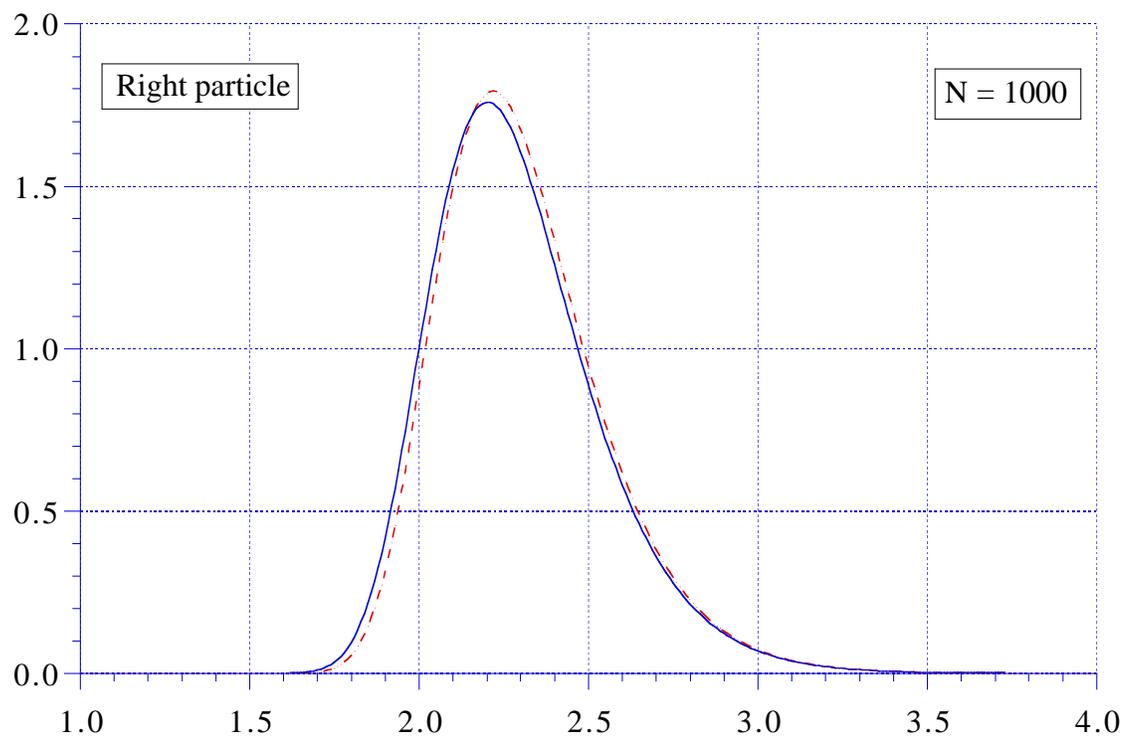
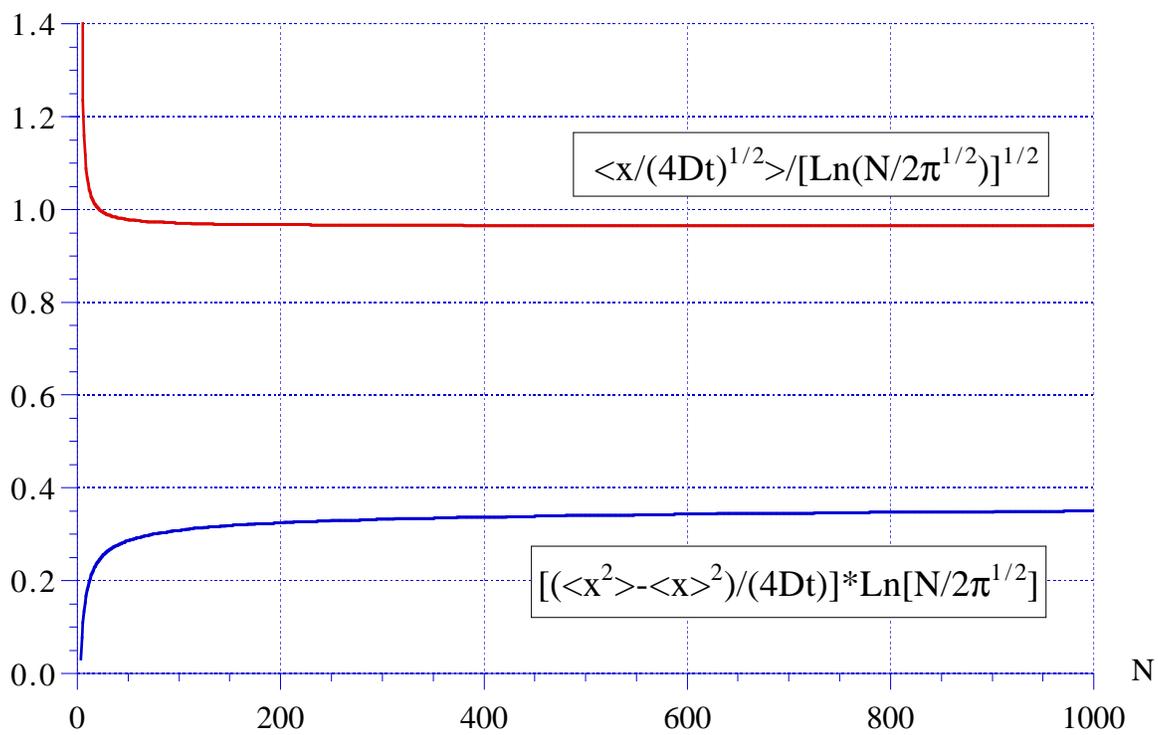
\end{document}